\documentclass[12pt,a4paper,fleqn]{article} 
\usepackage[fleqn]{amsmath} 
\usepackage{physics} 
\usepackage{cite} 
\usepackage[pdftex]{hyperref} 
\begin{document} 

\title{ 
\textbf{Plane-wave model of neutrino 
oscillations revisited}
}
\author{
\textit{Winfried A. Mitaroff} \footnote{ 
\quad \href{mailto:winfried.mitaroff@oeaw.ac.at}
{\texttt{winfried.mitaroff@oeaw.ac.at} } 
} \\
Institute of High Energy Physics, \\
Austrian Academy of Sciences, Vienna 
}
\date{10 December 2020} 


\maketitle
\thispagestyle{empty}

\begin{abstract} 

The phenomenology of massive neutrinos -- flavour mixing in the lepton sector 
causing oscillations between different neutrino-types along their propagation 
over macroscopic distances in vacuum -- aims at relating observable quantities 
(oscillation frequency or, equivalently, oscillation length) to the neutrino properties: 
mixing angles $\theta_{ij}$ and mass-squared differences $\Delta m_{ij}^2$. 

Calculation of the probabilities for a given neutrino-type either to survive or to mutate 
into another type, as functions of momentum $p$ and travelling distance $L$, 
are properly based on wave-packet models of varying complexity. 
Approximations neglecting subtle effects like decoherence result in the 
standard oscil\-lation formulae 
with terms proportional to $\sin^2(\Delta m_{ij}^2 L / 4 p)$. 

The same result may also be derived by a simple plane-wave model as shown in 
most textbooks. However, those approaches rely on unphysical a-priory assumptions: 
either ``equal energy'' or ``equal velocity'' or ``equal momentum'' in the phases of 
different mass eigenstates -- which are refuted elsewhere. 
In addition, some assume tacitly that interference occurs at time $t = L$. 

This study re-examines the plane-wave model. 
No unphysical assumption is necessary for deriving the standard formulae: 
a heuristic approach relies only on carefully defining interference at time $t = L / \beta$, 
and is justified by coherence arguments based in a qualitative way on wave-packets. 

\end{abstract} 


\section{Introduction}

A minimal extension of the Standard Model for non-zero mass neutrinos \footnote{ 
\, Present data do not exclude the possibility of the lowest mass being exactly zero. 
} 
requires  flavour-mixing in the lepton sector, which is described (analogously to 
quark mixing) by the $3 \times 3$ unitary Pontecorvo-Maki-Nakagawa-Sakata 
(PMNS) matrix $U$: 
\begin{equation} 
\label{pmns} 
\left( \begin{array} {c} 
\nu_e \\ \nu_{\mu} \\ \nu_{\tau} 
\end{array} \right) 
= U 
\left( \begin{array}{c} 
\nu_1 \\ \nu_2 \\ \nu_3 
\end{array} \right) 
\hspace{8mm} \mathrm{and} \hspace{8mm} 
\left( \begin{array} {c} 
\nu_1 \\ \nu_2 \\ \nu_3 
\end{array} \right) 
= U^{\dagger} 
\left( \begin{array}{c} 
\nu_e \\ \nu_{\mu} \\ \nu_{\tau} 
\end{array} \right) 
\end{equation} 
where $\ket{\nu_{\ell}}$ ($\ell = e, \mu, \tau$) is the weak eigenstate of a neutrino 
created or absorbed in some charged-current weak interaction (``flavour eigenstate''), 
and $\ket{\nu_i}$ ($i = 1, 2, 3$) is the eigenstate of the free-particle Hamiltonian 
(``mass eigenstate'') describing a neutrino's kinematic behaviour -- 
which is, however, not directly observable. 

After applying the unitary constraints and removing unphysical phases, 
the matrix elements of $U \equiv (U_{{\ell}i})$ and of its inverse
$U^{-1} = U^{\dagger} \equiv (U^{\ast}_{{\ell}i})^T$ 
can be parametrized by three rotation angles 
($\theta_{12}$, $\theta_{13}$, $\theta_{23}$) and one complex phase 
$\delta$, thus introducing the possibility of $CP$ violation in the lepton sector. 
Take care when a neutrino $\ket{\nu_{\ell}}$ couples as an adjoint spinor to the weak 
interaction vertex: its corresponding coupling factor (i.e. the matrix element 
$U_{{\ell}i}$ or $U_{{\ell}i}^{\ast}$) will appear as the complex conjugate \cite{Thomson}. 

A neutrino is produced weakly as a well-defined flavour state, but manifests itself as a 
\textit{coherent linear superposition} of the three mass eigenstates: 
\begin{equation} 
\label{psi0} 
\psi_{\ell} (0, 0) \equiv \ket{\nu_{\ell}} = \sum_{i=1}^3 U_{{\ell}i}^{\ast} \ket{\nu_i} 
\end{equation} 
\indent
Each $\ket{\nu_i}$'s momentum $p_i$ and energy $E_i = \sqrt{m_i^2 + p_i^2}$ 
are separately determined by energy-momentum conservation in the production process. 
This fact is exploited by precision experiments for measuring the neutrino masses 
\cite{Boehm}.\footnote{ 
\, So far, only upper limits can be derived from the effective masses squared \cite{Vogel}. 
} 
An example is the 2-body decay ${\pi}^+ \rightarrow  \nu_{\mu} \, {\mu}^+$ \cite{Boehm, Vogel}: 
if the $\pi$~mass, the $\mu$~mass and the $\mu$~momentum in the $\pi$ rest frame are known 
with sufficiently high precision, then the neutrino mass squared is kinematically determined. 
Such observation causes the superposition to collapse into one 
specific mass eigenstate $\ket{\nu_i}$ with probability $\left| U_{{\mu}i} \right|^2$; 
the measurements yield only an incoherently averaged muon-based 
\textit{effective mass squared} $\sum_{i=1}^3 \left| U_{{\mu}i} \right|^2 m_i^2$. 


\section{Plane-wave model} 
\label{planewave} 

In absence of such an observation, the wave function $\psi_{\ell}$ will propagate by evolving 
as \textit{coherently superposed plane-waves} along e.g.  the $x$-direction:\footnote{ 
\, Notwithstanding any uncertainties at production, free-particle propagation is always on-shell 
\cite{Akhmedov09}. 
Assuming vacuum only, no matter effects like MSW need to be taken into account. 
} 
\begin{equation} 
\label{psi2} 
\psi_{\ell} (t, x) = \sum_{i=1}^3 U_{{\ell}i}^{\ast} \ket{\nu_i} e^{- i \phi_i},
\hspace{6mm} \mathrm{phase} \: \phi_i = E_i \, t - p_i \, x
\end{equation} 
with different phases $\phi_i$ for each of its components $\ket{\nu_i}$. 
The interfering phases will steadily shift apart --
\textit{this dispersion is the origin of the oscillation}. 

Note that for any plane-wave, the phase velocity $\equiv E_i  /  p_i = 1 / {\beta}_i \ge 1$; 
the group velocity $\equiv \mathrm{d}E_i / \mathrm{d}p_i = p_i / E_i = {\beta}_i \le 1$ 
is equal to the particle's velocity in the lab frame. 
For a wave-packet, ${\beta}_i$ is the velocity of the packet's centre. 

The neutrino will eventually be detected at a distance $x = L$ by some 
charged-current weak interaction, and its absorbed flavour ${\ell}^{\ast}$ can be identified. 
Therefore, the mass eigenstates $\ket{\nu_i}$ of eq. (\ref{psi2}) have to be re-expressed 
in terms of flavour eigenstates $\ket{\nu_{{\ell}'}}$ while keeping into account the 
evolved individual phases $\phi_i$: 
\begin{equation} 
\label{psi3} 
\psi_{\ell} (t, x) = \sum_{{\ell}'=e}^{\tau} \left( \sum_{i=1}^3 
U_{{\ell}i}^{\ast} \, U_{{\ell}'i} e^{- i \phi_i} \right) \ket{\nu_{{\ell}'}} 
\end{equation} 
\indent
At this point the detection causes the wave function $\psi_{\ell}$ to collapse into a 
specific flavour state $\ket{\nu_{{\ell}^{\ast}}}$, 
and the probability of the original flavour $\ell$ to be observed in the detector as 
flavour ${\ell}^{\ast}$ (including survival if ${\ell}^{\ast} = \ell$) is 
\begin{equation} 
\label{prob} 
\mathcal{P}(\nu_{\ell} \rightarrow \nu_{{\ell}^{\ast}}) = 
\left| \braket{\nu_{{\ell}^{\ast}}}{\psi_{\ell} (t, L)} \right|^2 
\end{equation} 

Evaluating that in terms of PMNS matrix elements $U_{{\ell}i}$ and 
phases ${\phi}_i$ is straightforward, albeit involving some calculations. 
The results can be found in textbooks, e.g. \cite{Thomson}. 
As an example, the survival probability is given by 
\begin{equation} 
\label{pro3} 
\begin{array}{l l}
\mathcal{P}(\nu_{\ell} \rightarrow \nu_{\ell}) = 1 - \sum_{i=1}^3 4 \left| U_{{\ell}i} \right|^2 \left| U_{{\ell}j} \right|^2 
{\sin}^2 \left( \frac{{\phi}_i - {\phi}_j}{2} \right), & 
\mathrm{with} \: j = (i \bmod 3) + 1 
\end{array} 
\end{equation} 

\subsection{Two-flavour mixing} 
\label{twoflavour} 

The principal consequences of mixing (except $CP$-violating effects) are manifest 
by simply regarding only two neutrino flavours (say, $\nu_e$ and $\nu_{\mu})$ 
together with two mass eigenstates ($\nu_1$ and $\nu_2)$.\footnote{ 
\, These consequences hold, of course, also for  $\nu_e \leftrightarrow \nu_{\tau}$ 
and $\nu_{\mu} \leftrightarrow \nu_{\tau}$ mixing. 
} 
In this scenario, the PMNS matrix is reduced to a $2 \times 2$ orthogonal matrix 
with only one real parameter, the rotation angle $\theta = \theta_{12}$: 
\begin{equation} 
\label{2x2} 
\quad U = 
\left( \begin{array}{r r} 
  {\cos} \, {\theta} & {\sin} \, {\theta} \\ 
- {\sin} \, {\theta} & {\cos} \, {\theta} 
\end{array} \right) 
\hspace{8mm} \mathrm{and} \hspace{8mm} 
U^{-1}  = U^{T} = 
\left( \begin{array}{r r} 
{\cos} \, {\theta} & - {\sin} \, {\theta} \\ 
{\sin} \, {\theta} &   {\cos} \, {\theta} 
\end{array} \right) 
\end{equation} 
\indent 
The survival probability of $\ket{\nu_e}$ follows from eqs. (\ref{pro3}, \ref{2x2}) 
with $i = 1$, and the mutation probability $\ket{\nu_e} \rightarrow \ket{\nu_{\mu}}$ 
as 1-complement of that value: 
\begin{equation} 
\label{pro2} 
\begin{array}{l}
\mathcal{P}(\nu_e \rightarrow \nu_e) = 
1 - {\sin}^2(2 \, \theta) \, {\sin}^2\left( \frac{{\phi}_1 - {\phi}_2}{2} \right) \\[2mm]
\mathcal{P}(\nu_e \rightarrow \nu_{\mu}) = 1 - \mathcal{P}(\nu_e \rightarrow \nu_e) = 
{\sin}^2(2 \, \theta) \, {\sin}^2\left( \frac{{\phi}_1 - {\phi}_2}{2} \right) 
\end{array} 
\end{equation} 
\indent 
Defining the mean energy and mean momentum ($E$ and $p$, respectively), 
the energy and momentum differences ($\Delta E$ and $\Delta p$, respectively), 
and the velocity $\beta$ of a fictive particle moving with 
\textit{mean 4-momentum} = $(E, p, 0, 0)$: 
\begin{equation} 
\label{defs} 
\begin{array}{l l} 
E = (E_1 + E_2) / 2 & 
\qquad \Delta E = E_1 - E_2 \ne 0 \\[1mm]
p \: = \, (p_1 + p_2) / 2 & 
\qquad \Delta p \: = \, p_1 \, - p_2 \: \ne 0 \\[1mm]
\beta \, = \, p / E \neq ({\beta}_1 + {\beta}_2) / 2 \\[1mm]
\end{array} 
\end{equation} 
\noindent
hence, the difference between two phases in $\psi_e (t, x)$ is 
\begin{equation} 
\label{delta} 
\Delta \phi = 
{\phi}_1 - {\phi}_2 = \Delta E \, t - \Delta p \, x
\end{equation} 
\indent
The \textit{probabilities} of eqs. (\ref{pro2}) \textit{oscillate} both 
in space ($k \sim \Delta p$) and in time ($\omega \sim \Delta E$), 
with an amplitude depending on the mixing angle $\theta$: 
\begin{equation} 
\label{osci} 
\begin{array}{l} 
\mathcal{P}(\nu_e \rightarrow \nu_{\mu}) = 1 - \mathcal{P}(\nu_e \rightarrow \nu_e) = 
{\sin}^2(2 \, \theta) \, {\sin}^2 \left( \frac{\Delta E}{2} \, t - \frac{\Delta p}{2} \, x \right), 
\end{array} 
\end{equation} 
\noindent
but the oscillation in time is artificial, caused by the infinite  plane-waves 
instead of more adequate wave-packets. 
The oscillation in space is genuine and experimentally proven \cite{Thomson, Nakamura}. 
The task is how to relate $\Delta E$ and $\Delta p$ to the neutrinos' kinematic attributes, 
in particular to their masses $m_i$ and momenta $p_i$ (or energies $E_i$). 

Interference of phases $\phi_1$ and $\phi_2$ must occur at the 
\textit{same space-time point} $(t, x, 0, 0)$. 
Neglecting the sizes of production and detection regions w.r.t. their distance $L$,
the space point is at $x = L$ and is fixed. 
Defining the corresponding time $t$ is a more subtle question, 
the answer of which requires wave-packet arguments. 


\subsection{Heuristic approach} 
\label{correct}

Fixing the space point in eq. (\ref{delta}) at $x = L$,  the oscillation frequency $\Delta \phi$ 
of eqs. (\ref{pro2}) can easily be calculated without unphysical assumptions: 
\begin{equation} 
\label{eq1} 
\begin{array}{l} 
\Delta p = p_1 - p_2 = \frac{p_1^2 - p_2^2}{p_1 + p_2} = 
\frac{(E_1^2 - E_2^2) \, - \, (m_1^2 - m_2^2)}{2 \, p} = 
\frac{E \cdot \Delta E}{ p} - \frac{m_1^2 - m_2^2}{2 \, p} = 
\frac{\Delta E}{\beta} - \frac{\Delta m^2}{2 \, p} \\[3mm]
\Delta \phi = \Delta E \, t - \Delta p \, L = 
\Delta E \left( t - \frac{L}{\beta} \right) + \frac{\Delta m^2}{2 \, p} \, L, 
\hspace{8mm} \mathrm{with} \: \: \Delta m^2 = m_1^2 - m_2^2 
\end{array} 
\end{equation} 
\indent
Each neutrino mass eigenstate $\ket{\nu_i}$, or equivalently the centre of its wave-packet, 
arrives at a different time $t_i = L / {\beta}_i$ at $x = L$. 
Defining $T = L / \beta$ being the arrival time of the fictive particle of mean 4-momentum 
in eq. (\ref{defs}), and assuming w.l.o.g. masses $m_1 < m_2$, 
then velocities ${\beta}_1 > \beta > {\beta}_2$ and arrival times $t_1 < T < t_2$. 

Interference $\ket{\nu_1} \leftrightarrow \ket{\nu_2}$ requires the phase difference 
${\phi}_1 - {\phi}_2$ to be observed at a common time $t$,\footnote{ 
\, The importance of the ``same space-time point'' condition for interference can best be 
illustrated by violating it, e.g. by taking the phases ${\phi}_i$ at the different arrival 
times $t_i$: $E_i \, t_i = E_i \, L / {\beta}_i = E_i^2 \, L / p_i = (m_i^2 + p_i^2) \, L / p_i 
\: \Rightarrow \: {\phi}_i = E_i \, t_i - p_i \, L = m_i^2 \, L / p_i \: \Rightarrow \: \Delta \phi = 
{\phi}_1 - {\phi}_2 = (m_1^2 / p_1 - m_2^2 / p_2) \, L \approx \Delta m^2 \, L / p$, which is 
obviously wrong -- it disagrees by the factor 2 w.r.t. the correct value of eq. (\ref{eq2}). 
} 
at which a ``snapshot'' of the interference pattern can be taken. 
Therefore $\ket{\nu_1}$ and $\ket{\nu_2}$ must be described as wave-packets 
of finite size in $x$ which partly overlap at $x = L$. 
Since $t_1$ and $t_2$ are the arrival times of the packets' centres, 
it is reasonable to fix interference at time $t = T = L / {\beta}$ \cite{Boehm}. 
With eqs. (\ref{osci}, \ref{eq1}) 
\begin{eqnarray} 
\label{eq2} 
& & \hspace{-6mm} 
t - \frac{L}{\beta} = 0 \qquad \Longrightarrow \qquad 
\Delta \phi = \frac{\Delta m^2}{2 \, p} \, L \\
\label{eq3} 
& & \hspace{-6mm} 
\mathcal{P}(\nu_e \rightarrow \nu_{\mu}) = 1 - \mathcal{P}(\nu_e \rightarrow \nu_e) = 
{\sin}^2(2 \, \theta) \, {\sin}^2 \left( \frac{\Delta m^2}{4 \, p} \, L \right) 
\end{eqnarray} 
\indent
The oscillation length $L_{osc}$ is found by equating the argument to $\pm \, \pi$ 
(conversion to conventional units done with $\hbar c = 0.19733$ GeV$\cdot$fm): 
\begin{equation} 
\label{leng} 
L_{osc} = 4 \pi \, \frac{p}{\left| \Delta m^2 \right|} \approx 
4 \pi \, \frac{E}{\left| \Delta m^2 \right|} = \, 2.48 \cdot 
\frac{E / \mathrm{GeV}}{\left| \Delta m^2 \right| / \mathrm{eV}^2} \: \mathrm{km} 
\end{equation} 
\indent 
Eqs. (\ref{eq3}) and (\ref{leng}) agree with the standard oscillation formulae. 

\subsection{Wave-packet aspects} 
\label{packet}

In sec. \ref{correct}, fixing interference at $t = T = L / {\beta}$ may look arbitrary. 
In fact, any time $t$ within the \textit{coherence time interval} $\Delta t $ at which the 
wave-packets overlap at $x = L$ could be chosen as well. 
Which are the consequences if $t \ne T$ ? 

Wave-packets are characterized by their finite size ${\sigma}_x$ in $x$. 
The tail of the faster $\ket{\nu_1}$ arrives at $t_{1T} = t_1 + {\sigma}_x / 2 {\beta}_1$, and 
the front of the slower $\ket{\nu_2}$ arrives at $t_{2F} = t_2 - {\sigma}_x / 2 {\beta}_2$ 
(remember $t_1 < t_2$). 
In order to be able to interfere, the wave-packets must overlap 
\cite{Giunti03, Akhmedov09, Akhmedov12}. 
Hence, the coherence time interval is defined by 
\begin{equation} 
\begin{array}{l} 
\Delta t = t_{1T} - t_{2F} \approx \frac{{\sigma}_x}{\beta} - (t_2 - t_1) > 0 
\end{array} 
\end{equation} 
and shrinks from originally ${\sigma}_x / \beta$ (full overlap at production) 
as the wave-packets get increasingly staggered in the course of propagation.\footnote{ 
\, Decoherence by complete separation ($t_2 - t_1 >  {\sigma}_x / \beta$) 
may occur at cosmic distances. 
}

For $t$ within this interval, and re-writing eq. (\ref{eq1}): 
\begin{equation} 
\label{eq4} 
\begin{array}{l} 
\left| t - T \right| < \Delta t \le \frac{{\sigma}_x}{\beta} \\[3mm]
\Delta \phi = \frac{\Delta m^2}{2 \, p} \, L + \Delta {\phi}', 
\hspace{12mm} \mathrm{with} \: \: 
\left| \Delta {\phi}' \right| = \left| \Delta E \right| \cdot \left| t - T \right| < 
\frac{\left| \Delta E \right|}{\beta} \, {\sigma}_x 
\end{array} 
\end{equation} 
being an \textit{additional phase shift} caused by the finite packet size. 

Further analyses by a realistic wave-packet model \cite{Akhmedov09} identify 
\textit{coherence conditions} for possible interference,\footnote{ 
\, The coherence condition $\left| \Delta E \right| \ll {\sigma}_E$ \cite{Akhmedov09}, 
together with Heisenberg's approximate uncertainty relation 
${\sigma}_x \cdot {\sigma}_p \approx 1$ and the dispersion relation 
${\sigma}_E / {\sigma}_p \approx \mathrm{d} E / \mathrm{d} p = p / E = \beta$, 
yield eq. (\ref{sigx}). 
} 
which subsequently constrain ${\sigma}_x$ to be 
\begin{equation} 
\label{sigx} 
\begin{array}{l l} 
{\sigma}_x \ll \frac{\beta}{\left| \Delta E \right|} 
\end{array} 
\end{equation} 
\indent 
Thus, the term $\Delta {\phi}'$ can be neglected, and eq. (\ref{eq4}) reproduces 
eq. (\ref{eq2}). 
This is, in retrospect, a justification for the heuristic ansatz of sec. \ref{correct}. 

\subsection{Textbook approaches} 
\label{textbook}

The conventional plane-wave approaches are based on unphysical assumptions:  
they mani\-festly violate energy-momentum conservation,\footnote{ 
\, Energy-momentum conservation is essential for the measurement of neutrino masses. 
}
and also violate Lorentz invariance \cite{Giunti01}. 
They are outlined below for stimulating criticism. \\[-3mm]

\begin{itemize} 
\item 
\textbf{Equal energy} ($E_1 = E_2$) 

Starting from eq. (\ref{eq1}) and setting $\Delta E = 0$ trivially yields 
$\Delta \phi$ of  eq. (\ref{eq2}). \\[-3mm]

\item
\textbf{Equal velocity} (${\beta}_1 = {\beta}_2$) 

Hidden as an exercise in \cite{Thomson}: 
starting from eq. (\ref{eq1}), setting ${\beta}_1 = {\beta}_2 = \beta$ and fixing  
$t = L / \beta$ yields $\Delta \phi$ of  eq. (\ref{eq2}). 
This approach misses the fact that assuming ${\beta}_1 = {\beta}_2$ is not necessary: 
it is sufficient to properly define $\beta$ as in sec. \ref{correct} above. \\[-3mm]

\item
\textbf{Equal momentum} ($p_1 = p_2$) 

A somewhat confused derivation, using the approximation $m_i \ll p_i \approx p$: \\

\begin{math} 
\begin{array}{l l} 
E_i = \sqrt{p_i^2 + m_i^2} \approx p_i + \frac{m_i^2}{2 \, p_i} \qquad \Longrightarrow & 
\Delta E \approx \Delta p + \frac{\Delta m^2}{2 \, p} \\[3mm]
\Delta \phi = \Delta E \, t - \Delta p \, L \approx 
\Delta p \left( t - L \right) + \frac{\Delta m^2}{2 \, p} \, t 
\end{array} 
\end{math} \\[2mm]

Setting $\Delta p = 0$ yields $\Delta \phi \approx \frac{\Delta m^2}{2 \, p} \, t$, 
thereafter tacitly  fixing $t = T_0 = L$ yields $\Delta \phi$ of  eq. (\ref{eq2}). 
But assuming $\Delta p = 0$ was not necessary: fixing $t$ would suffice. 

A closer look shows that $T_0$ is the time of arrival at $x = L$ of a fictive zero-mass particle, 
and $T_0 < t_1 < t_2$. Interference requires the slower wave-packet $\ket{{\nu}_2}$ 
to have a big enough size for its front having arrived at $L$ already at time $t = T_0$. 
\\[-3mm]
\end{itemize} 


\section{Summary}

A thourogh review of neutrino oscillations, covering both theoretical and experimental 
aspects, is given in \cite{Nakamura} with exhaustive references. 
There is general consent that a proper treatment can only be based on sophisticated 
 wave-packet models. 
 Many such models exist, but are too complex for usual textbooks. 

Resorting to the simpler plane-wave model, conventional approaches, however, 
 rely on unphysical a-priory assumptions (outlined in sec. \ref{textbook}) 
which are vigorously refuted in several theoretical papers, 
e.g. \cite{Giunti03, Akhmedov09, Akhmedov12}. 
But they still dominate most textbooks, 
sometimes with caveats and a reference to wave-packets \cite{Thomson}. 

Sec. \ref{correct} presents a non-conventional heuristic approach, 
based on the same plane-wave model, albeit without relying on any  
of those unphysical assumptions. The key is to pay attention 
how to define a time at which interference takes place \cite{Boehm}. 
The results agree with the standard formulae shown in the textbooks. 

Sec. \ref{packet} gives further justification for this heuristic approach, 
using coherence arguments based qualitatively on wave-packets \cite{Akhmedov09}. 

Hopefully this study will contribute to a better understanding of the plane-wave model, 
and may inspire the authors of future textbooks. 

\subsection*{Acknowledgement}

Thanks are due to \textit{Walter Grimus} (University of Vienna) for 
helpful comments and suggestions, and for a careful reading of the manuscript. 


\small

\normalsize

\end{document}